\pdfoutput=1

\documentclass[11pt,twoside,a4paper,cmspaper,final,collab]{cms-tdr}
\def\svnVersion{298927}\def\cmsCernNo{2013-037}\def\cmsCernDate{\today}\def\cmsMessage{Published in the Journal of High Energy Physics as \href{http://dx.doi.org/10.1007/JHEP06(2015)121}{\doi{10.1007/JHEP06(2015)121.}}}

\begin{document}\cmsNoteHeader{B2G-14-004}

\hyphenation{had-ron-i-za-tion}
\hyphenation{cal-or-i-me-ter}
\hyphenation{de-vices}
\RCS$Revision: 288965 $
\RCS$HeadURL: svn+ssh://svn.cern.ch/reps/tdr2/papers/B2G-14-004/trunk/B2G-14-004.tex $
\RCS$Id: B2G-14-004.tex 288965 2015-05-15 01:59:41Z alverson $

\newcommand{\procLumi}{19.7\fbinv}
\providecommand{\mt}{\ensuremath{M_{\mathrm{T}}}\xspace}

\cmsNoteHeader{B2G-14-004}

\title{Search for the production of dark matter in association with top-quark pairs in the single-lepton final state in proton-proton collisions at  $\sqrt{s} = 8$\TeV}

\date{\today}

\abstract{
A search is presented for particle dark matter produced in association with a pair of top quarks in pp collisions at a centre-of-mass energy of $\sqrt{s} = 8$\TeV. The data were collected with the CMS detector at the LHC and correspond to an integrated luminosity of 19.7\fbinv. This search requires the presence of one lepton, multiple jets, and large missing transverse energy. No excess of events is found above the SM expectation, and upper limits are derived on the production cross section. Interpreting the findings in the context of a scalar contact interaction between fermionic dark matter particles and top quarks, lower limits on the interaction scale are set. These limits are also interpreted in terms of the dark matter-nucleon scattering cross sections for the spin-independent scalar operator and they complement direct searches for dark matter particles in the low mass region.
}

\hypersetup{%
pdfauthor={CMS Collaboration},%
pdftitle={Search for the production of dark matter in association with top-quark pairs in the single-lepton final state in proton-proton collisions at  sqrt(s) = 8 TeV},
pdfsubject={CMS},%
pdfkeywords={CMS, physics, dark matter}}

\maketitle

\section{Introduction}
\label{sec:intro}

Dark matter (DM) is estimated to account for about 23\% of the total mass of the universe, and to be five times more abundant than the known baryonic matter. While the existence of DM is inferred from astrophysical observations, there is very little information about its nature or how it interacts with ordinary matter.

In this paper, we consider a simplified scenario~\cite{Maverick,JHEP10081,PhysRevD.82.116010} in which DM has a particle explanation and, in particular, there is only one new Dirac fermion related to DM within the energy reach of the LHC. The fermion interacts with quarks via a four-fermion contact interaction, which can be described by an effective field theory (EFT) Lagrangian:
\begin{equation}
L_{\text{int}} = \sum_{q} \sum_{i} C_{q\,i}~ \bigl(\overline{q}\Gamma_{i}^{q}{\mathrm{q}}\bigr)\bigl(\overline{\chi}\Gamma_{i}^{\chi}\chi\bigr),
\label{eft}
\end{equation}
where $C$ represents the coupling constant, which usually depends on the scale of the interaction ($M_{*}$). The operator $\Gamma$ describes the type of the interaction, including scalar ($\Gamma=1$), pseudoscalar ($\Gamma=\gamma^{5}$), vector ($\Gamma=\gamma^{\mu}$), axial vector ($\Gamma=\gamma^{\mu}\gamma^{5}$), and tensor interactions ($\Gamma=\sigma^{\mu\nu}$).  The exact value of the constant $C$ depends on the particular type of the interaction.

This scenario can lead to the production of DM particles in association with a hard parton, a photon, or a W or Z boson.  The first two production modes are usually referred to as monojets~\cite{Maverick,PhysRevD.82.116010,PhysRevD.85.056011,PhysRevD.86.015010,PhysRevD.84.095013} and monophotons~\cite{PhysRevD.85.056011}, respectively. Recent monojet results from the ATLAS~\cite{atlas7tev} and CMS~\cite{CMS:rwa} Collaborations have placed lower limits on $M_{*}$ for some typical couplings in Eq.~(\ref{eft}). The ATLAS Collaboration~\cite{PhysRevLett.112.041802} has also searched for DM particles in events with a hadronically decaying W or Z boson. Assuming a DM particle with a mass of 100\GeV, the excluded interaction scales are below about 60\GeV~\cite{PhysRevLett.112.041802}, 1040\GeV~\cite{CMS:rwa}, 1010\GeV~\cite{CMS:rwa}, and 2400~\cite{PhysRevLett.112.041802}\GeV for scalar, vector, axial-vector, and tensor interactions, respectively, and the excluded scale is below 410\GeV~\cite{CMS:rwa} for a scalar interaction between DM particles and gluons.

The exclusion limit for a scalar interaction between DM particles and quarks is the least stringent among all the interaction types that have been probed. In
this interaction the coupling strength is proportional to the mass of the quark:
\begin{equation}
L_{\text{int}} = \frac{m_{\PQq}}{M^{3}_{*}} \overline{q}{q}\overline{\chi}\chi.
\label{d1}
\end{equation}

As a consequence, couplings to light quarks are suppressed. A recent paper~\cite{PhysRevD.88.063510} suggested that the sensitivity to the scalar interaction can be improved by searching in final states with third-generation quarks. It has also been noted that the inclusion of heavy quark loops in the calculation of monojet production ~\cite{Haisch:2012kf} increases the expected sensitivity.

\begin{figure}[tbh]
 \begin{center}
\includegraphics[width=0.35\textwidth]{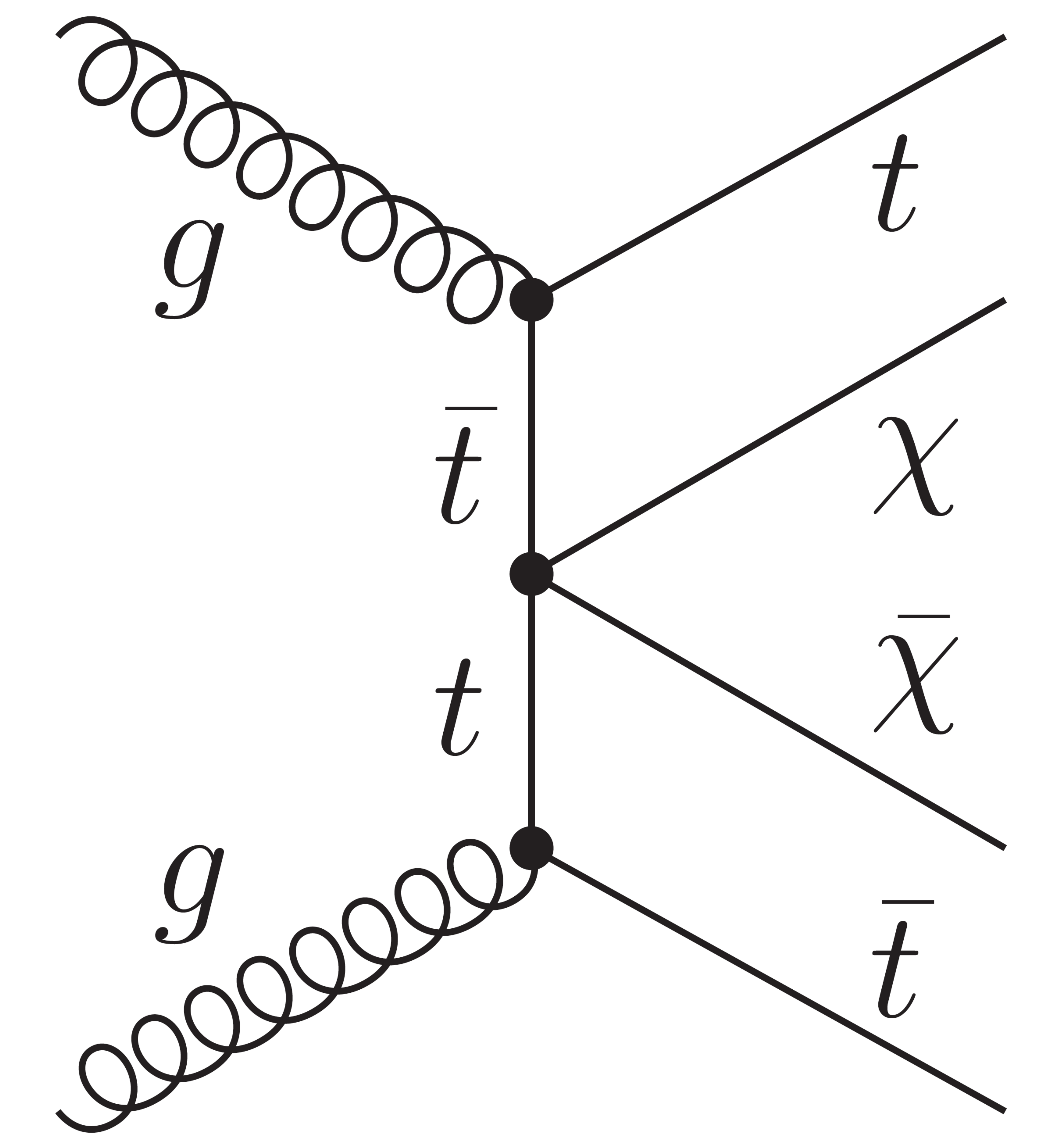}
   \caption{Dominant diagram contributing to the production of DM particles in association with top quarks at the LHC.}
   \label{fig:ggtottxx}
 \end{center}
\end{figure}

In this paper, we report on a search for the production of DM particles in association with a pair of top quarks, and consider only the scalar interaction. The ATLAS Collaboration has recently searched for DM particles in association with heavy quarks~\cite{ATLAS_heavyQuarks}, placing more stringent limits on the scalar interaction between DM particles and quarks than the mono-W/Z search~\cite{PhysRevLett.112.041802}. Assuming a DM particle with a mass of 100\GeV, the excluded interaction scale is 120\GeV for scalar interaction between top quarks and DM particles. Figure~\ref{fig:ggtottxx} shows the dominant diagram for this production at the LHC. In this paper we focus our search on events with one lepton (electron or muon) in the final state.

\section{The CMS Detector}
\label{sec:CMS}

The central feature of the CMS apparatus is a superconducting solenoid of 6\unit{m} internal diameter, providing a magnetic field of 3.8\unit{T}.  Within the field volume are a silicon pixel and strip tracker, a lead tungstate crystal electromagnetic calorimeter (ECAL), and a brass and scintillator hadron calorimeter (HCAL), each composed of a barrel and two endcap sections. Muons are measured in gas-ionization detectors embedded in the steel flux-return yoke outside the solenoid. Extensive forward calorimetry complements the coverage provided by the barrel and endcap detectors. A more detailed description of the CMS detector, together with a definition of the coordinate system used and the relevant kinematic variables, can be found in Ref.~\cite{Chatrchyan:2008zzk}.

\section{Data and simulated samples}
The data used in this search were recorded with the CMS detector at the LHC at $\sqrt{s}= 8$\TeV, and correspond to an integrated luminosity of $\procLumi$.  The data were collected using single-electron and single-muon triggers, with transverse momentum (\pt) thresholds of 27 and 24\GeV, respectively.  The efficiencies of these triggers in data and simulation are compared, measured using a tag-and-probe method~\cite{tagAndProbe}, and correction factors are applied to the simulation.

DM signals are generated with \MADGRAPH v5.1.5.11~\cite{madgraph} leading order (LO) matrix element generator using the CTEQ6L1 parton distribution functions (PDF)~\cite{Pumplin:2002vw}. The dominant standard model (SM) background processes for this search are \ttbar{}+jets,  $\ttbar+\gamma/\PW/\Z$, \PW+jets, single top quark, diboson (WW, WZ, and ZZ) and Drell--Yan events. All of these backgrounds except single top quark and WW events, are generated with the \MADGRAPH using CTEQ6L1 PDF.  The top-quark \pt distributions in the \ttbar{}+jet sample generated from \MADGRAPH are reweighted to match the CMS measurements, following the method described in Ref.~\cite{topPt}. Single top quark processes are generated with the next-to-LO (NLO) generator \POWHEG v1.0 using the CTEQ6M PDF~\cite{Pumplin:2002vw}.  The WW background is generated with the \PYTHIA v6.424~\cite{pythia}. All events generated with \MADGRAPH are matched to the \PYTHIA~\cite{pythia} parton shower description.  All events are passed through the detailed simulation of the CMS detector based on {\sc GEANT4} v9.4~\cite{Agostinelli:2002hh}.

The cross sections of \ttbar{}+jets~\cite{Czakon:2013goa} and $\PW/\Z$+jets~\cite{Quackenbush2013209} backgrounds are calculated at next-to-NLO. Other backgrounds are calculated at NLO. The single top quark cross section is taken from Ref.~\cite{Kidonakis:2012db}, the $\ttbar+\Z$ cross section from Ref.~\cite{ttWttZ_NLO}, the $\ttbar+\PW$ cross section from Ref.~\cite{ttW_NLO}, the $\ttbar+\gamma$ cross section from Ref.~\cite{PhysRevD.83.074013} and the diboson cross sections are from Ref.~\cite{VV_LHC}.

Additional minimum bias events in the same LHC bunch crossing (pileup) are added to all simulated events, with a distribution in number matching that observed in data.

\section{Object reconstruction}
A particle-flow (PF) based event reconstruction~\cite{PFT-09-001,PFT-10-001} is used by CMS, which takes into account information from all subdetectors, including charged-particle tracks from the tracking system and deposited energy from the ECAL and HCAL. Given this information, all particles in the event are classified into mutually exclusive categories: electrons, muons, photons, charged hadrons, and neutral hadrons. Primary vertices are reconstructed using a deterministic annealing filter algorithm~\cite{Chabanat:865587}, with the event primary vertex defined as the vertex with the largest sum of the squares of the \pt of the tracks associated with that vertex.

Electron candidates are reconstructed from energy clusters in the ECAL matched with tracks ~\cite{epjcs10052-006-0175-5}. The electron trajectory in the tracker volume is reconstructed with a Gaussian sum filter~\cite{gsf} algorithm that takes into account the possible emission of bremsstrahlung photons in the silicon tracker. The electron momentum is then determined from the combination of ECAL and tracker measurements. Electrons are identified by placing requirements on the ECAL shower shape, the matching between the tracker and the ECAL, the relative energy fraction deposited in HCAL and ECAL, the transverse and longitudinal impact parameters of the tracker track with respect to the event primary vertex, photon conversion rejection, and the isolation variable $R^{\Pe}_{\text{Iso}}$. The isolation variable is defined as the ratio to the electron transverse momentum, of the sum of \pt of all other PF candidates reconstructed in a cone of radius $\Delta R=\sqrt{\smash[b]{(\Delta\eta)^{2} + (\Delta\phi)^{2}}} = 0.3$ around the electron candidate, where $\eta$ is the pseudorapidity and $\phi$ is the azimuthal angle. The \pt sum in the isolation cone is corrected for the contributions of pileup interactions on an event-by-event basis. Isolated electrons satisfy $R^{\Pe}_{\text{Iso}}<0.1$. The electron is required not to be in the transition region between the barrel and the endcap ECAL ($1.44<\abs{\eta}<1.57$) because the reconstruction of an electron object in this region is not optimal~\cite{epjcs10052-006-0175-5}. After all these requirements, electrons are selected if they satisfy $\pt > 30$\GeV and $\abs{\eta} <2.5$.

Muon candidates are reconstructed by combining tracks from the tracker and muon system~\cite{1748-0221-7-10-P10002}, resulting in ``global-muon tracks''. The PF muons are selected among reconstructed muon track candidates by imposing minimal requirements on the track components in the muon system and taking into account matching with small energy deposits in the calorimeters~\cite{PFT-09-001,PFT-10-001}. Muons from cosmic rays and from light hadrons that decay in flight, or from b hadrons, and hadrons misidentified as muons are suppressed by applying requirements on the quality of the global-muon fit, the number of hits in the muon detector and in the tracker, the transverse and longitudinal impact parameters of the tracker track with respect to the event primary vertex, and the isolation variable. The muon isolation variable ($R^{\mu}_{\text{Iso}}$) is defined in a similar manner to that for electrons, but with a cone of radius $\Delta R=0.4$. Isolated muons must satisfy $R^{\mu}_{\text{Iso}}<0.12$. After all these requirements, muons are selected if they satisfy $\pt > 30$\GeV and $\abs{\eta} <2.1$.

Both electron and muon identification efficiencies are measured via the tag-and-probe technique using inclusive samples of $\Z\to\ell^{+}\ell^{-}$ events from data and simulation. Correction factors are used to account for the difference in performance of the lepton identification between data and simulation.

Jets are reconstructed from PF candidates that are clustered with the anti-\kt algorithm~\cite{Cacciari:2008gp} with a distance parameter of 0.5, using the \FASTJET package~\cite{Cacciari:2011ma}. Jet energy scale corrections obtained from data and simulation are applied to account for the response function of the combined calorimetry to hadronic showers and pileup effects~\cite{Soyez:2012hv,Perloff:2012wpa}.  The jet \pt resolution in simulation is adjusted to match that measured in data~\cite{JECpaper}. Jet candidates are required to have $\pt>30$\GeV and $\abs{\eta} < 4.0$, and to satisfy a very loose set of quality criteria~\cite{JECpaper}. The combined secondary vertex (CSV) b-tagging algorithm~\cite{BTAGpaper} is used to identify jets from the hadronization of b quarks. The CSV algorithm exploits the large impact parameters and probable presence of a displaced vertex which are common in b-quark-initiated jets. This information is combined in a likelihood discriminant providing a continuous output between 0 and 1. In this search, a selected jet is considered to be b-tagged if it has a CSV discriminant value greater than 0.679 and $\abs{\eta} <2.4$. The b-tagging efficiency is approximately 70\% (20\%) for jets originating from a b (c) quark and the mistagging probability for jets originating from light quarks or gluons is approximately 2\%. An event-by-event correction factor is applied to simulated events to account for the difference in performance of the b-tagging between data and simulation~\cite{bTagSF}.

Missing transverse energy (\MET) is measured as the magnitude of the vectorial \pt sum of all PF candidates, taking into account the jet energy corrections.

\section{Event selection}
\label{selection}
In semileptonic $\ttbar$ decays, two b quarks and two light quarks are produced. Therefore most of the selected signal events contain at least four jets. However, we set the requirement to be three or more rather than four  or more identified jets in an event, since this is found to improve the search sensitivity by 10\%.  In addition, we require at least one b-tagged jet (``b jet'') in the event, and only one identified isolated lepton.

Signal events usually have larger $\MET$ than the backgrounds because of two DM particles, neither of which leave any energy in the detector. Events are therefore required to have $\MET > 160$\GeV. These selection criteria are referred to as the ``preselection''. After preselection, the dominant backgrounds are from \ttbar and W+jets production. Other backgrounds include single top, Drell--Yan and diboson production. The QCD multijet contribution to the background is negligible because of the requirements of a high-\pt isolated lepton, large $\MET$, and a b-tagged jet.

To improve the search sensitivity, we further select events with $\MET> 320$\GeV. The remaining \PW+jets and most \ttbar backgrounds contain
a single leptonically decaying W boson. The transverse mass, defined as $\mt\equiv \sqrt{\smash[b]{2\MET \pt^\ell(1-\cos(\Delta\phi))}}$, where $\pt^\ell$ is the transverse momentum of the lepton and $\Delta\phi$ is the opening angle in azimuth between the lepton and $\ptvecmiss$ vector, is constrained kinematically to $\mt<M_{\PW}$ for the on-shell W boson decay in the \ttbar and \PW+jets events. For signal events, off-shell W boson decays, and \ttbar dilepton decay channel, \mt can exceed $M_{\PW}$. Therefore a requirement of $\mt> 160$\GeV is applied to increase the discrimination of the background relative to the signal.

\begin{figure}[tbh]
 \begin{center}
   \includegraphics[width=0.28\textwidth]{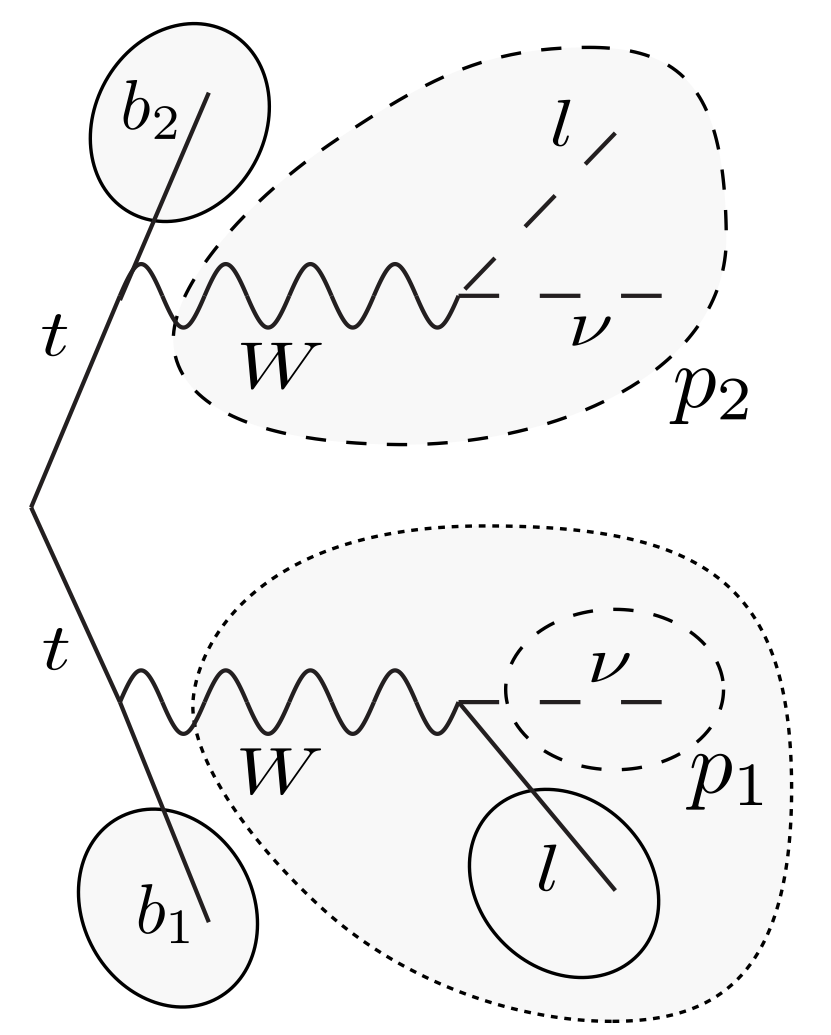}
   \caption{Schematic of a dileptonic \ttbar event where only one lepton is reconstructed~\cite{MT2W}. This represents the dominant type of \ttbar background to this search. The momentum of the W boson that decays to an unreconstructed lepton is indicated by $p_2$, and the momentum of the neutrino from the decay of the other W boson is indicated by $p_1$. The same notation is used in Eq.~(\ref{eq:mt2w}).}
   \label{fig:mt2w}
 \end{center}
\end{figure}

The dominant background with large \mt arises from dileptonic \ttbar events where one of the leptons is unobserved, illustrated in Fig.~\ref{fig:mt2w}. The $M_{\mathrm{T2}}^{\PW}$ variable~\cite{MT2W} is exploited to further reduce this type of background.  This variable is defined as the minimal ``parent'' particle mass compatible with all the transverse momenta and mass-shell constraints, assuming two identical parent particles, each of mass $m_\mathrm{y}$, decaying to bW:
\begin{equation}
M_{\mathrm{T2}}^{\PW}  =  \min \left( m_{\mathrm{y}} \text{ consistent with: } \left\{
\begin{aligned}
&\vec{p}_{1}^{\mathrm{T}} + \vec{p}_{2}^{\mathrm{T}} = \ptvecmiss, p_{1}^{2}=0, (p_{1} + p_{\ell})^{2} = p_{2}^{2} = M^{2}_{\PW}, \\
&(p_{1}+p_{\ell}+p_{\mathrm{b1}})^{2} = (p_{2} + p_{\mathrm{b2}})^{2} = m^{2}_{\mathrm{y}}
\end{aligned}
\right\} \right),
\label{eq:mt2w}
\end{equation}

where the momentum of the W boson that decays to an unreconstructed lepton is indicated by $p_2$, and the momentum of the neutrino from the decay of the other W boson is indicated by $p_1$. In particular, the intermediate W bosons are assumed to be on-shell, thus adding more kinematic information to suppress dileptonic \ttbar events where one lepton is lost. In \ttbar events, the $M_{\mathrm{T2}}^{\PW}$ distribution has a kinematic end-point at the top-quark mass, assuming perfect measurements with the detector. By contrast, this is not the case for signal events where two additional DM particles are present. The calculation of $M_{\mathrm{T2}}^{\PW}$ requires that at least two b jets be identified and be paired correctly to the lepton. When only one b jet is selected, each of the first three remaining highest \pt jets is considered as the second b jet. When two or more b jets are selected, all the b jets in the event are used. The $M_{\mathrm{T2}}^{\PW}$ value is then calculated for all possible jet-lepton combinations and the minimum value is taken as the event discriminant. We select events with $M_{\mathrm{T2}}^{\PW} > 200$\GeV.

In addition, the jets and the $\ptvecmiss$ tend to be more separated in $\phi$ in signal events than in \ttbar background. We therefore require the minimum opening angle in $\phi$ between each of the first two leading jets and $\ptvecmiss$ to be larger than 1.2. In summary, the signal region (SR) for our search is $\MET>320$\GeV, $\mt>160$\GeV, $M_{\mathrm{T2}}^{\PW} > 200$\GeV and min $\Delta\phi(j_{1,2},\ptvecmiss) > 1.2$.  These selection criteria are optimized based on the expected significance for DM masses between 1 and 1000\GeV.

\begin{figure}[phbt]
 \begin{center}
   \includegraphics[width=0.45\textwidth]{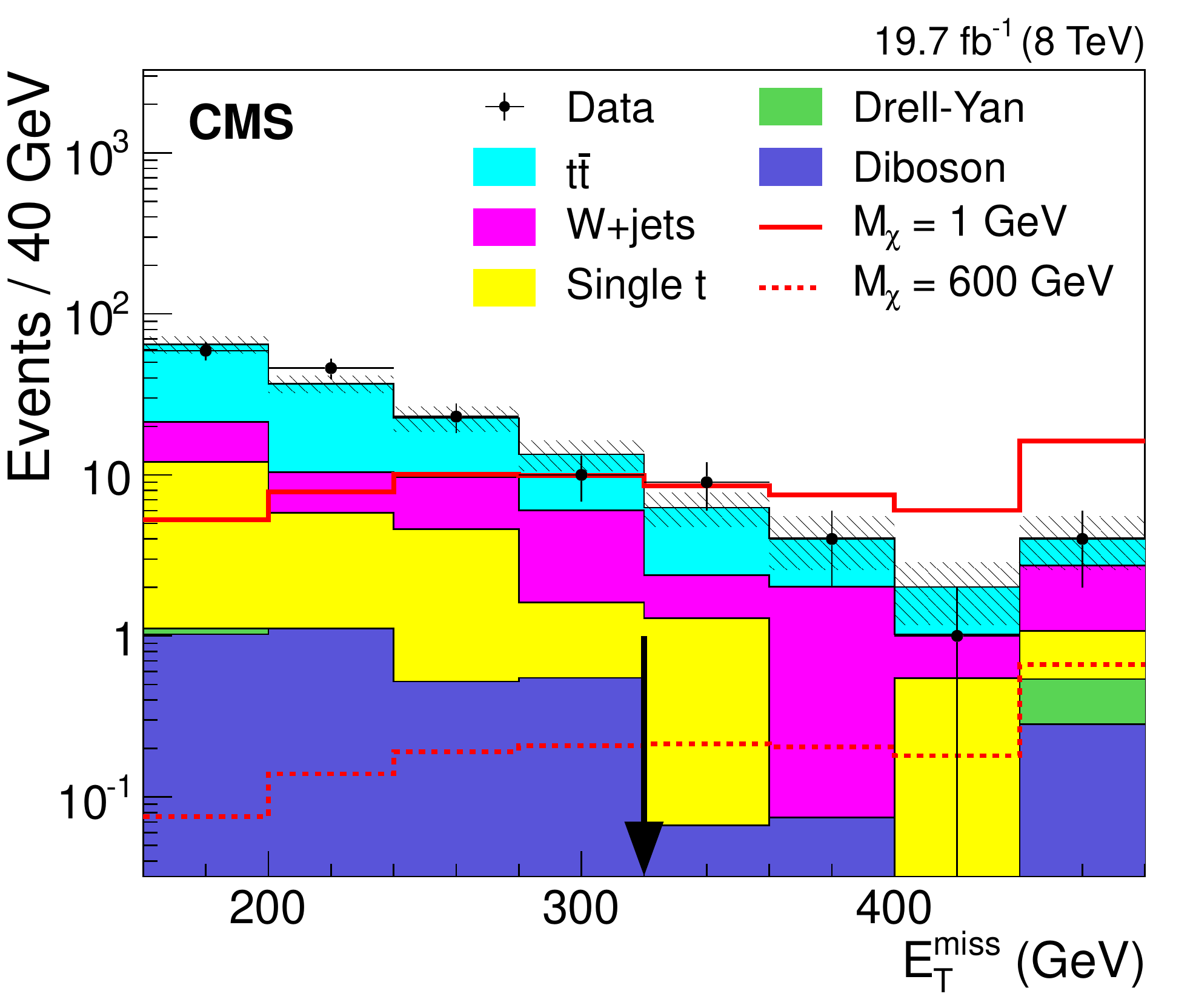}
   \includegraphics[width=0.45\textwidth]{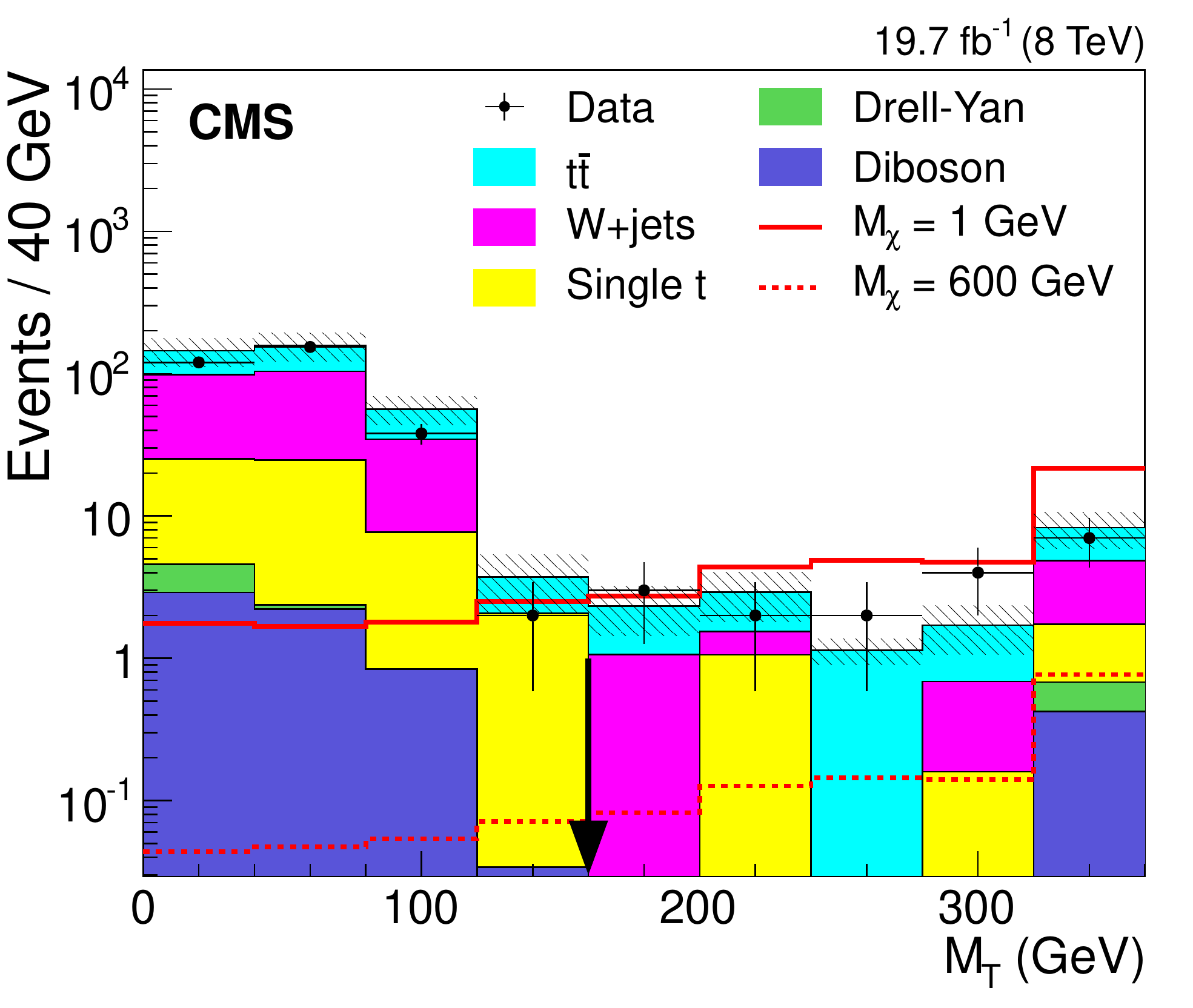}
   \includegraphics[width=0.45\textwidth]{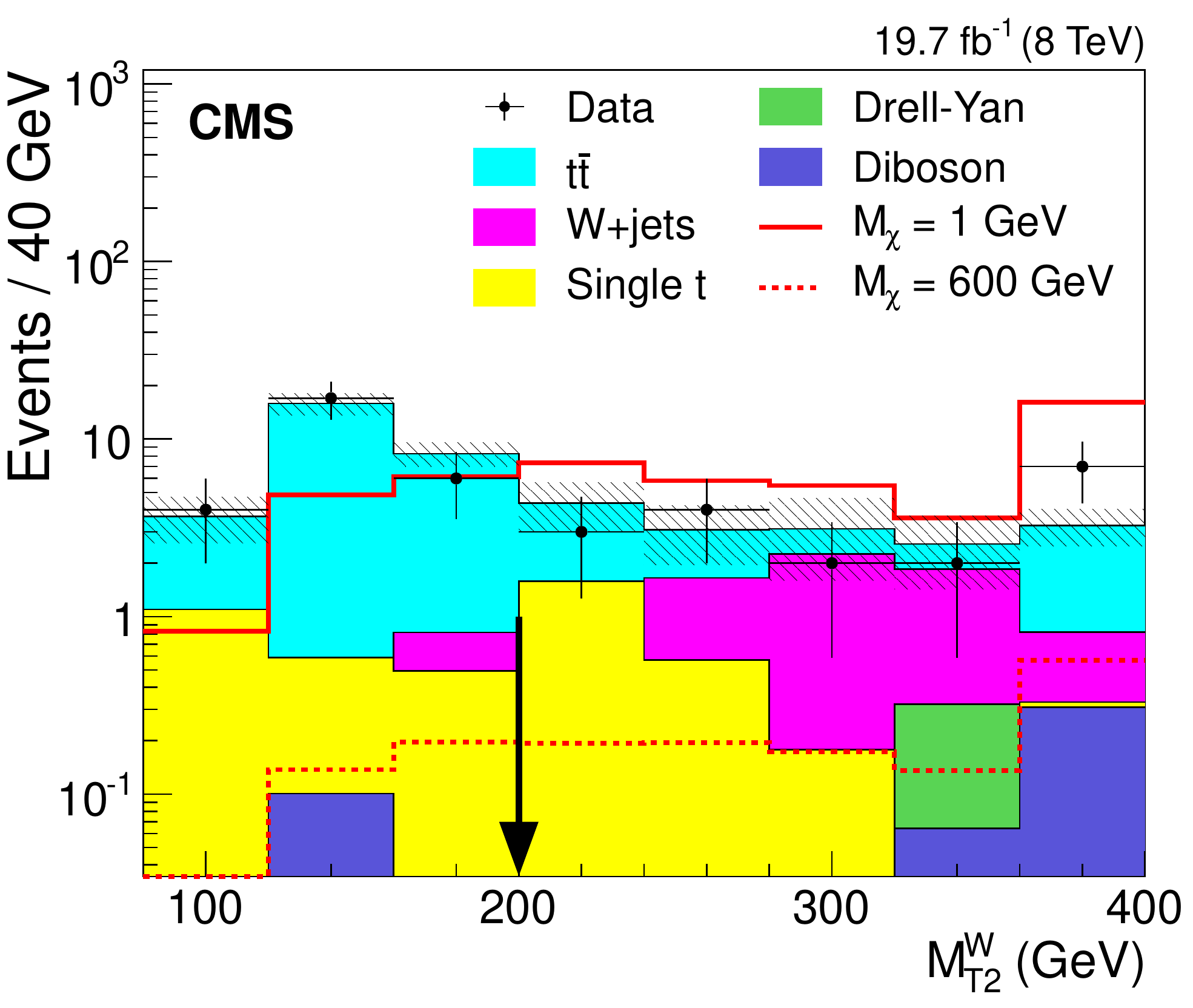}
   \includegraphics[width=0.45\textwidth]{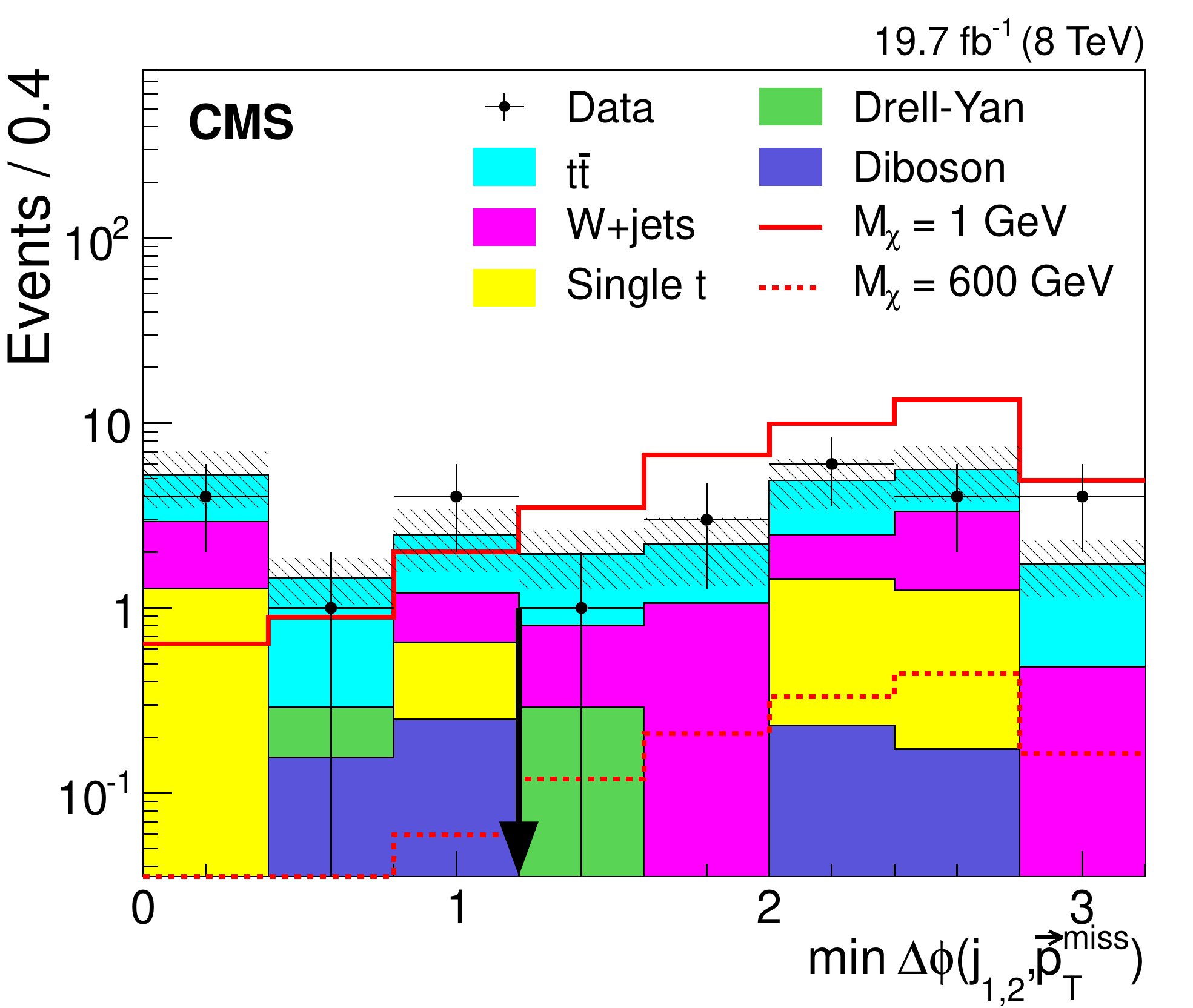}
  \caption{Distributions of $\MET$, $\mt$, $M_{\mathrm{T2}}^{\PW}$, and min $\Delta\phi(j_{1,2},\ptvecmiss)$ after applying SFs for \ttbar{}+jets and \PW+jets backgrounds, as described in section~\ref{sec:background}. Each distribution is plotted after applying all other selections, which are indicated by the arrows on the relevant distributions. Two simulated DM signals with mass $M_{\chi}$ of 1 and 600\GeV and an interaction scale $M_{*}$ of 100\GeV are included for comparison. The hatched region represents the total uncertainty in the background prediction. The last bin of the \MET, \mt and $M_{\mathrm{T2}}^{\PW}$ distributions includes the overflow. The horizontal bar on each data point indicates the width of the bin.}
  \label{fig:cut26}
\end{center}
\end{figure}

Figure~\ref{fig:cut26} shows the distributions of \MET, $\mt$, $M_{\mathrm{T2}}^{\PW}$, and min $\Delta\phi(j_{1,2},\ptvecmiss)$ after applying all other selections except the one plotted, indicating their power of discrimination between signal and background. In these distributions, the \ttbar{}+jets and \PW+jets backgrounds have been adjusted by the scale factors (SF), as described in section~\ref{sec:background}.

\section{Background estimation}
\label{sec:background}

Standard model backgrounds are estimated from simulation, with data-to-simulation SFs applied
to the dominant backgrounds from \ttbar{}+jets and \PW+jets.

\begin{figure}[p!hbt]
 \begin{center}
   \includegraphics[width=0.45\textwidth]{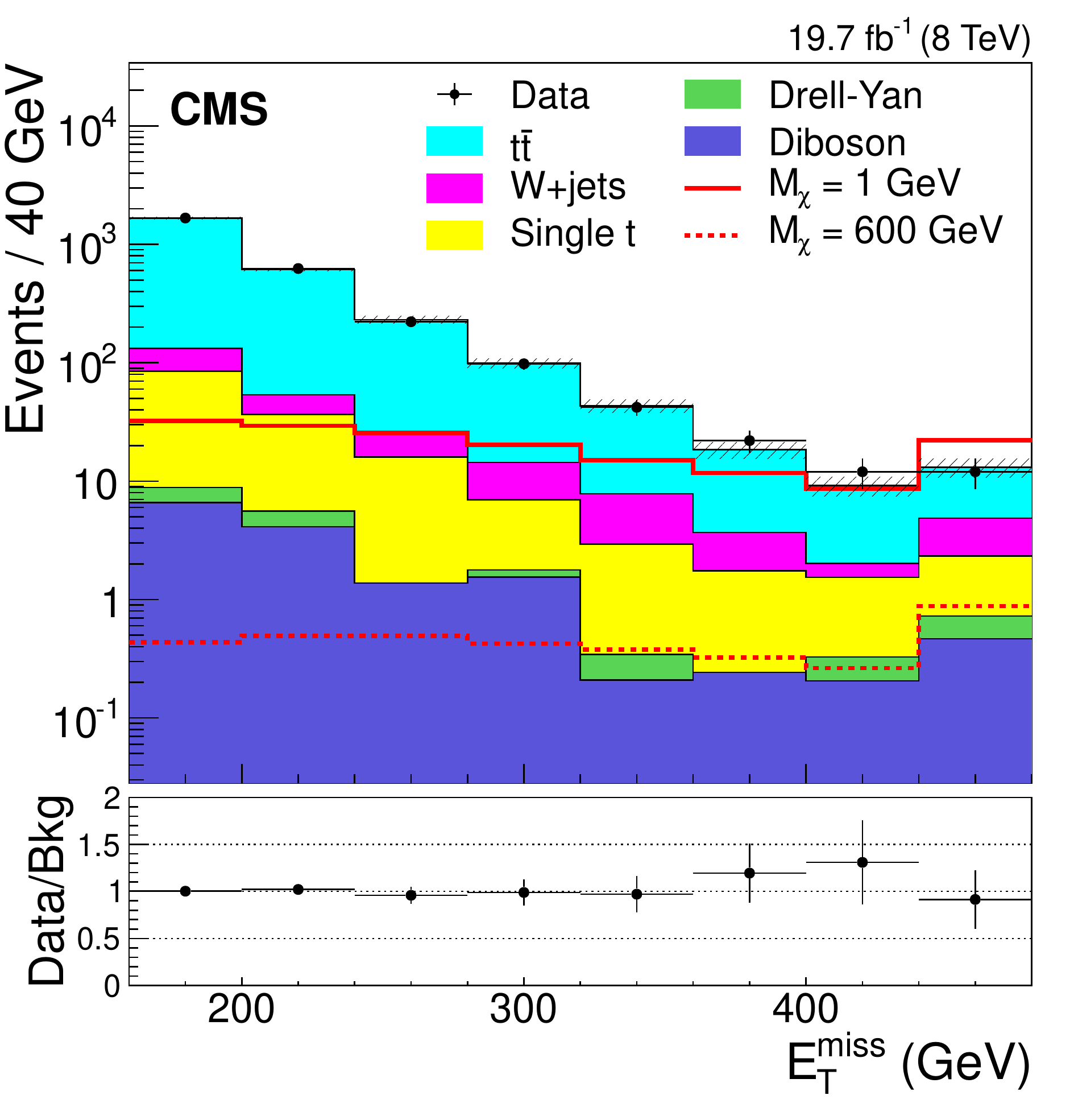}
   \includegraphics[width=0.45\textwidth]{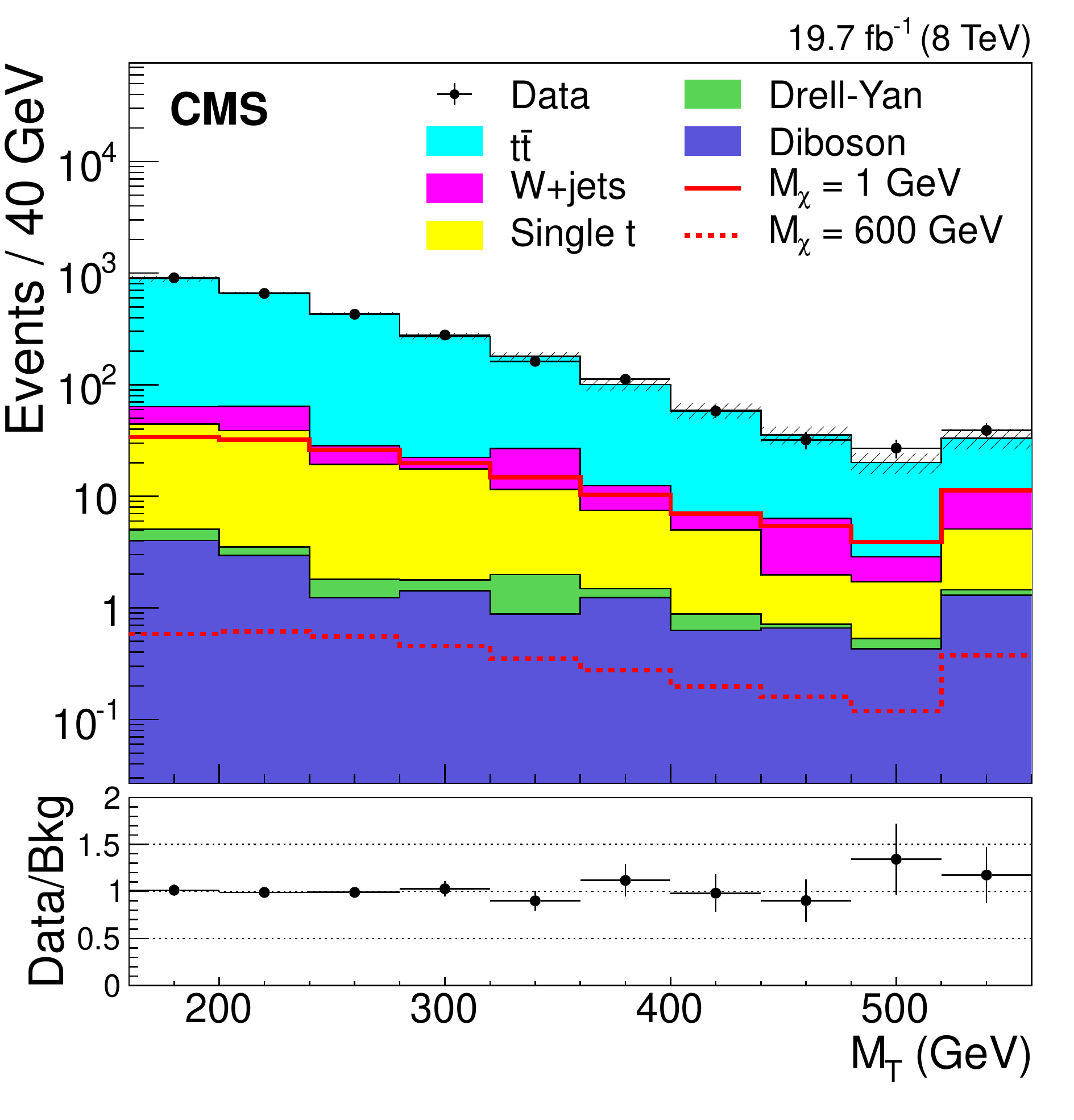}
   \includegraphics[width=0.45\textwidth]{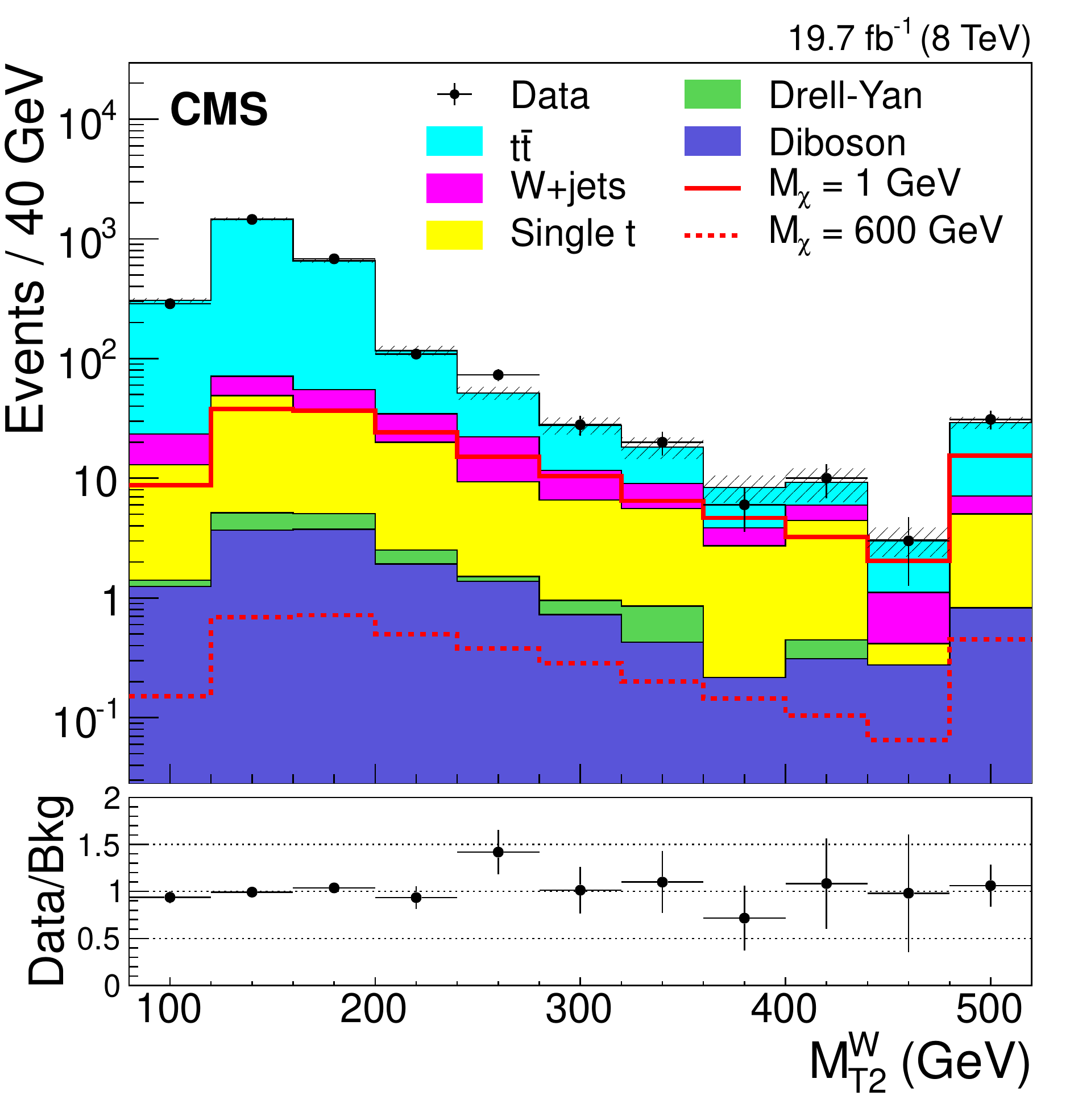}
   \includegraphics[width=0.45\textwidth]{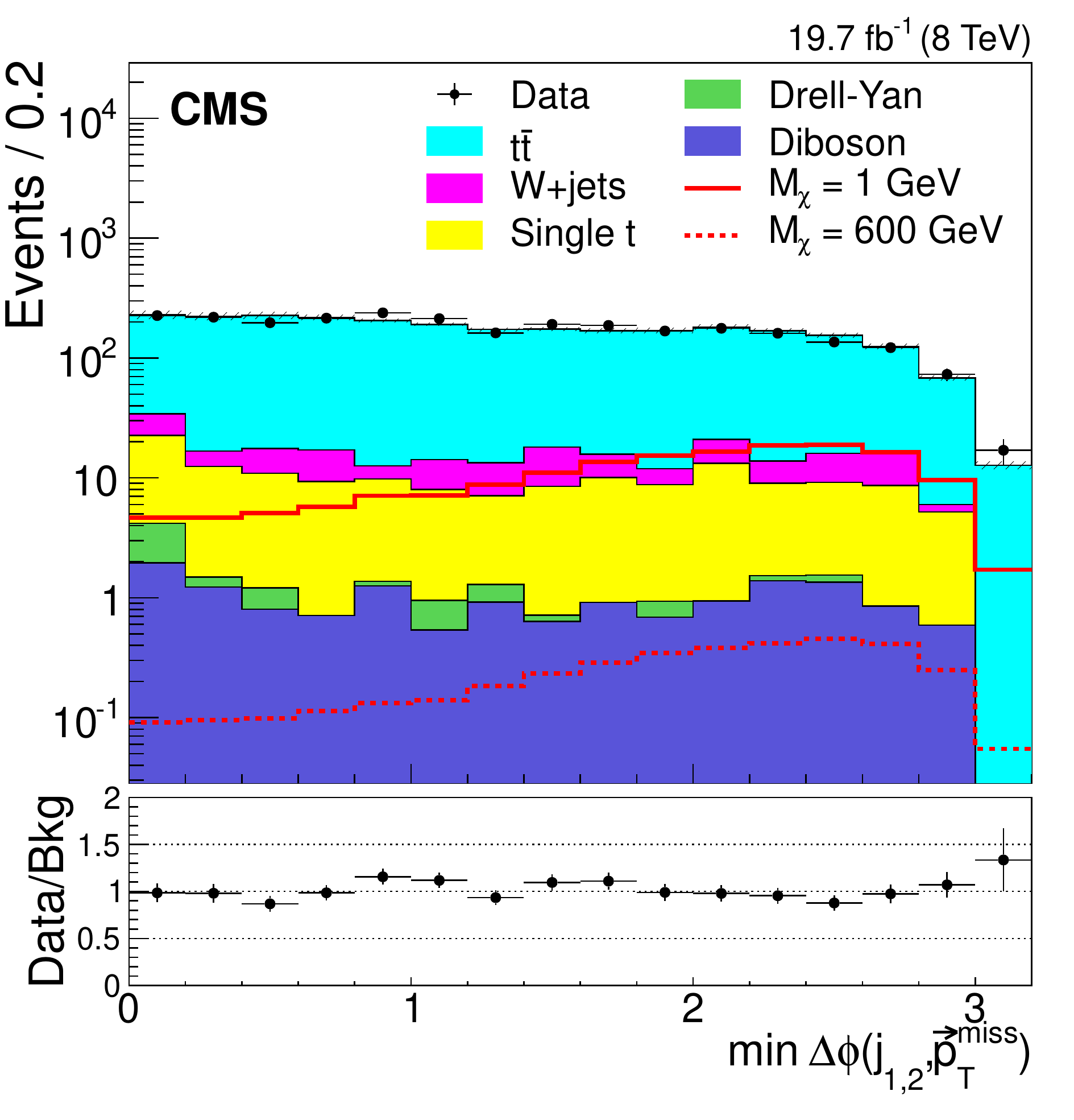}
   \caption{Distributions of $\MET$, $\mt$, $M_{\mathrm{T2}}^{\PW}$, and min $\Delta\phi(j_{1,2},\ptvecmiss)$ in CR1 after applying the SFs for \ttbar{}+jets and \PW+jets backgrounds, as described in section~\ref{sec:background}. Two simulated DM signals with mass $M_{\chi}$ of 1 and 600\GeV and an interaction scale $M_{*}$ of 100\GeV are included for comparison. The hatched region represents the total uncertainty in the background prediction. The error bars on the data-to-background ratio take into account both the statistical uncertainty in data and the total uncertainty in the background prediction. The last bin of the \MET, \mt, and $M_{\mathrm{T2}}^{\PW}$ distributions includes the overflow. The horizontal bar on each data point indicates the width of the bin.}
   \label{fig:afterFitCR1}
 \end{center}
\end{figure}

\begin{figure}[phbt]
 \begin{center}
  \includegraphics[width=0.45\textwidth]{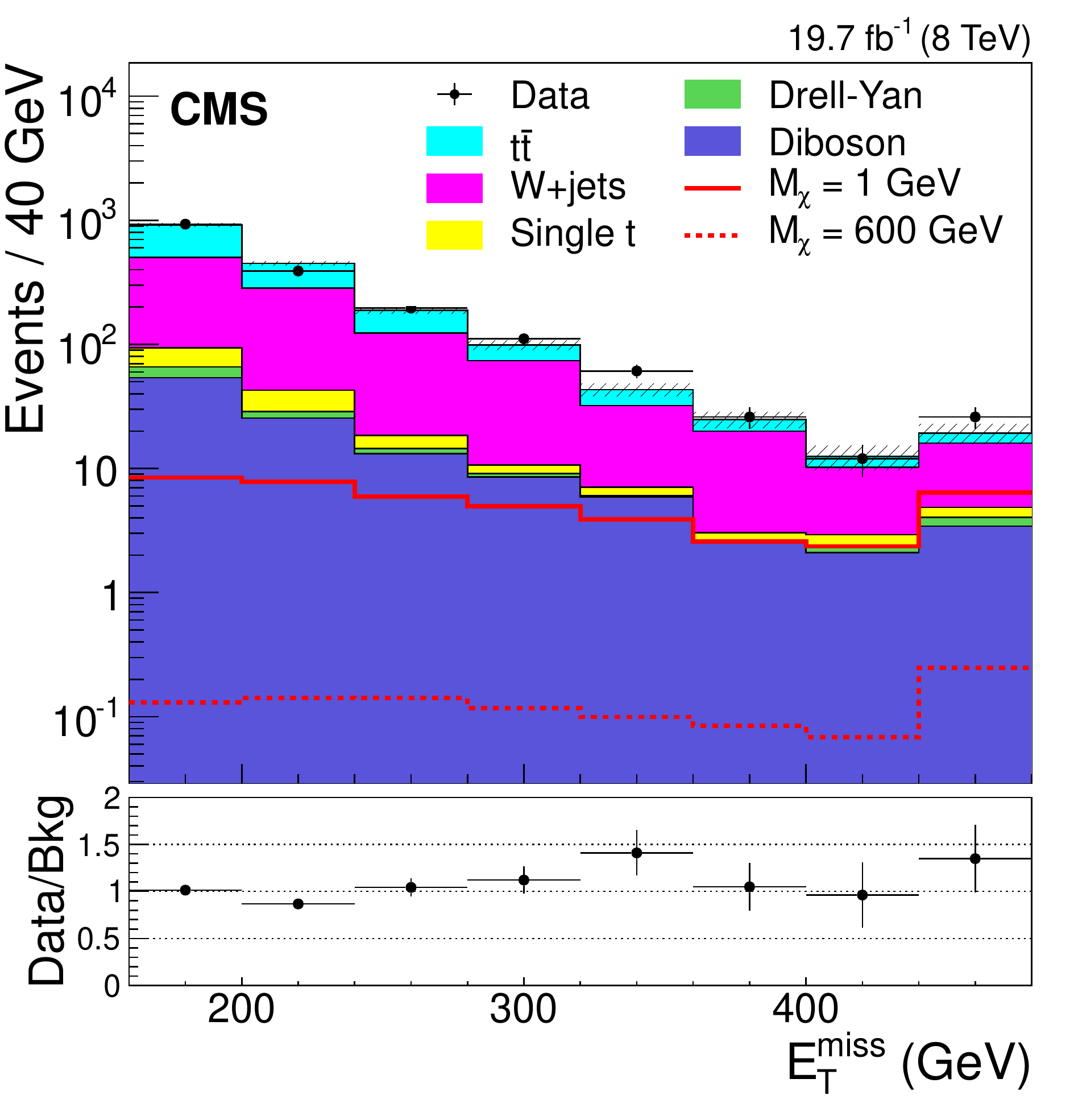}
  \includegraphics[width=0.45\textwidth]{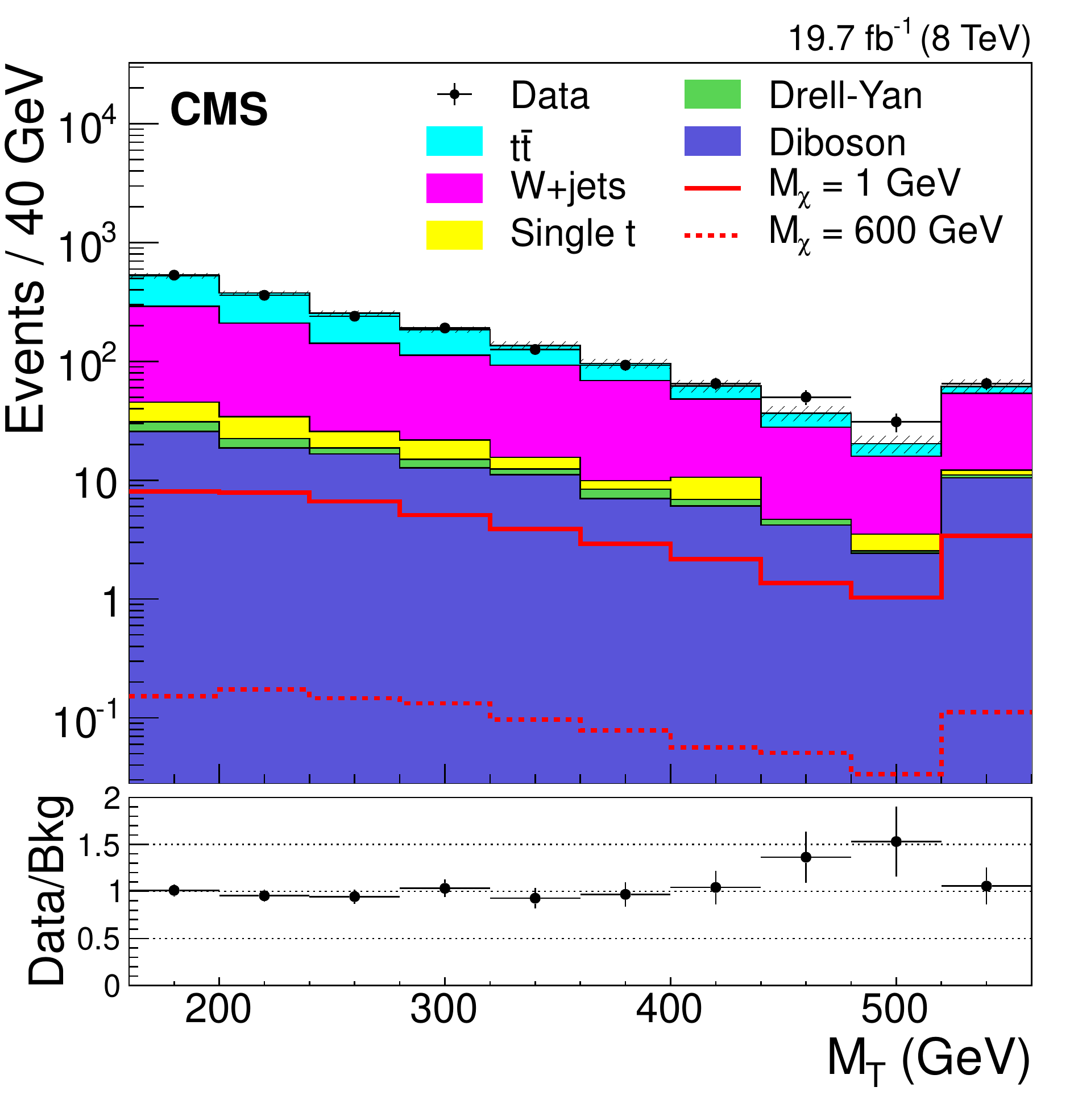}
  \includegraphics[width=0.45\textwidth]{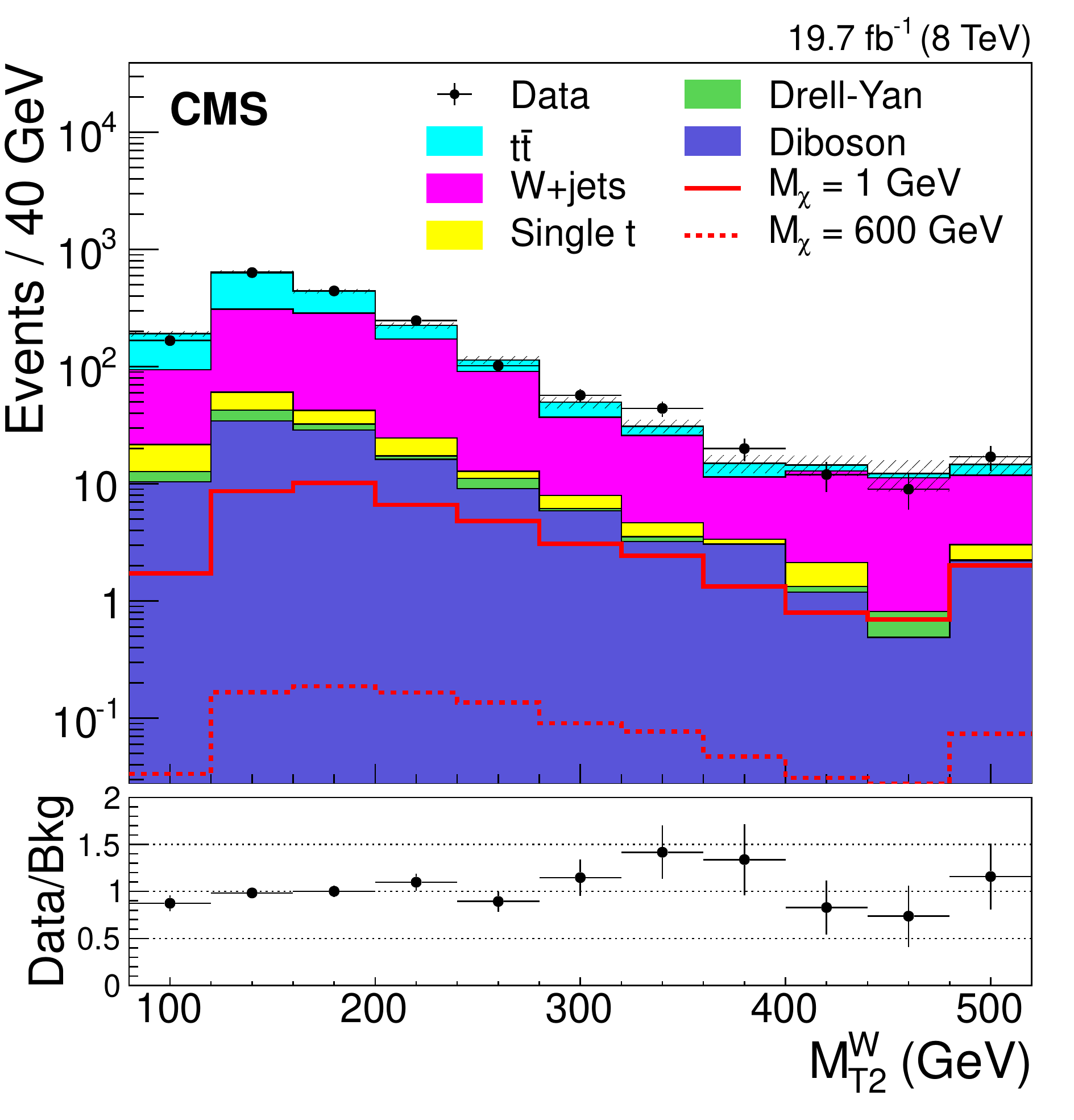}
  \includegraphics[width=0.45\textwidth]{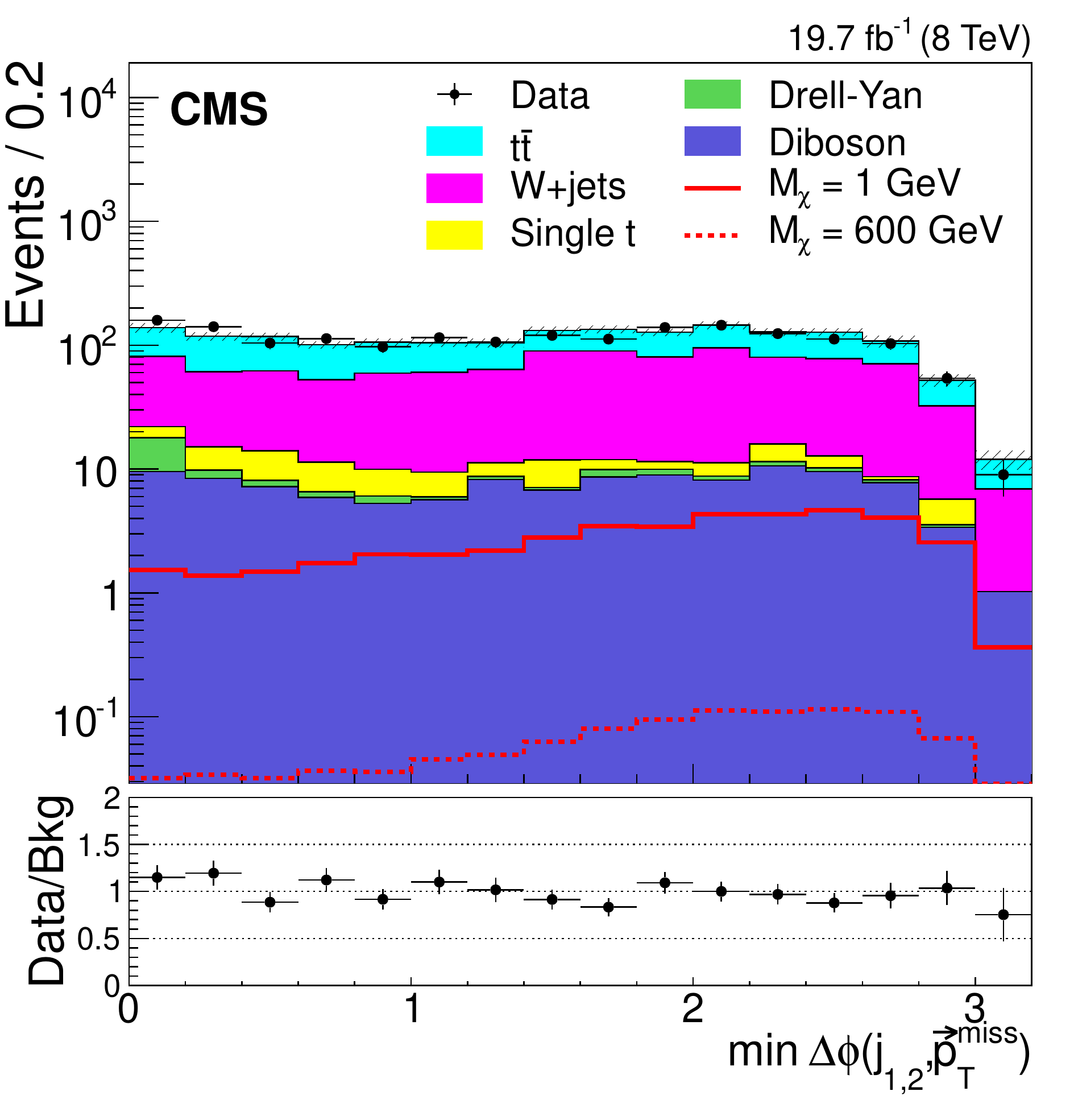}
    \caption{Distributions of $\MET$, $\mt$, $M_{\mathrm{T2}}^{\PW}$, and min $\Delta\phi(j_{1,2},\ptvecmiss)$ in CR2 after applying the SFs for \ttbar{}+jets and \PW+jets backgrounds, as described in section~\ref{sec:background}. Two simulated DM signals with mass $M_{\chi}$ of 1 and 600\GeV and an interaction scale $M_{*}$ of 100\GeV are included for comparison. The hatched region represents the total uncertainty in the background prediction. The error bars on the data-to-background ratio take into account both the statistical uncertainty in data and the total uncertainty in the background prediction. The last bin of the \MET, \mt, and $M_{\mathrm{T2}}^{\PW}$ distributions includes the overflow. The horizontal bar on each data point indicates the width of the bin.}
   \label{fig:CR2}
 \end{center}
\end{figure}

Two control regions (CR) are defined to extract these SFs. One is the preselection with the additional requirement of $\mt>160\GeV$ (CR1). The sample in CR1 is dominated by \ttbar{}+jets background. The other (CR2) is defined the same way as CR1 except that no jet should satisfies the b-tag requirement, resulting in a sample enriched in \PW+jets events. The subdominant backgrounds are subtracted from the distributions observed in data in order to obtain a data sample that has only \ttbar{}+jets and \PW+jets background contributions. The \ttbar{}+jets and \PW+jets SFs are then obtained by matching simultaneously to data the $\mt$ distribution in CR1 and the $\MET$ distribution in CR2. The obtained SFs for \ttbar{}+jets and \PW+jets are $1.11\pm0.02$\stat and $1.26\pm0.06$\stat, respectively.  These SFs are propagated to the SR to estimate the background. The level of DM signal contamination in the two CRs is estimated to be small and therefore has negligible impact on the background estimation in the SR. Figures~\ref{fig:afterFitCR1} and~\ref{fig:CR2} show the distributions of $\MET$, $\mt$, $M_{\mathrm{T2}}^{\PW}$, and min $\Delta\phi(j_{1,2},\ptvecmiss)$ with the SFs applied in CR1 and CR2, respectively. The data are in good agreement with expectations from SM background.

\section{Systematic uncertainties}
\label{sys}

The normalization and shape of the distributions used to establish a possible DM signal are subject both to experimental and theoretical uncertainties.

The data-to-simulation SFs for \ttbar{}+jets and \PW+jets are extracted from the CRs, as described in the previous section. For the background estimation, the use of SFs largely removes the uncertainties from the integrated luminosity, lepton identification and trigger efficiencies, and from cross sections of the two backgrounds. Other systematic uncertainties can be constrained by refitting the data in the CRs, as described in the following.

The \ttbar{}+jets and \PW+jets SFs are obtained from CRs in which other backgrounds are present as well. We conservatively assign a 50\% uncertainty for other backgrounds to account for possible missing higher order terms as well as mismodelling of kinematic properties from the simulation. This uncertainty results in a change of 5\% and 9\% for the \ttbar{}+jets and \PW+jets SFs, respectively. Propagating these changes to the SR, the impact on the total background prediction is found to be 10\%.

The stability of the SFs is checked through changes in the definitions of the CRs. These include tightening
the $\MET$ requirement or applying selections on $M_{\mathrm{T2}}^{\PW}$, and min $\Delta\phi(j_{1,2},\ptvecmiss)$.
An uncertainty of 40\% for the \PW+jets SF is assigned from these CR tests. No significant change is observed in the
SF for \ttbar{}+jets.

The \pt distributions of top quarks in the \ttbar{}+jets simulation is reweighted to match the data. The reweighting uncertainty is estimated by changing the nominal reweighting factor to unity or to the square of the reweighting factor, resulting in a change of $\pm$14\% for the \ttbar{}+jets SF and only negligible impact on the \PW+jet SF. Propagating these SFs to the SR, a systematic uncertainty of 10\% is estimated for the \ttbar{}+jets background prediction from the reweighting. The stability of the \ttbar{}+jets background prediction is also checked by varying the \MADGRAPH factorization and renormalization scale parameters, or the scale parameter for the matrix element and parton shower matching, by a factor of two. The resulting predictions are consistent with the nominal \ttbar{}+jets background prediction.

The remaining dominant experimental systematic uncertainties are from corrections in jet energy scale and resolution. Correction factors are separately varied by $\pm$1 standard deviation and $\MET$ is recalculated accordingly. These changes in the jet energy scale and resolution correction factors contribute uncertainties of 4\% and 3\% in the estimate of the background, respectively. The uncertainties in the background yield due to b-tagging correction factors are estimated to be 1.0\% and 1.8\% for heavy-flavour and light-flavour jets, respectively. The uncertainty in the pileup model contributes an uncertainty of 2.0\% in the background estimate.

The theoretical uncertainty related to the choice of the PDF set is evaluated by reweighting the background samples using three PDF sets: CT10~\cite{CT10}, MWST2008~\cite{MSTW2008}, and NNPDF2.3 ~\cite{NNPDF2.3}, following the PDF4LHC recommendation~\cite{PDF4LHC,Botje:2011sn}. For each PDF set, an uncertainty band is derived from the different error PDF sets, including the uncertainties due to the strong coupling constant $\alpha_{\mathrm{S}}$. The envelope of these three error bands is taken as the PDF uncertainty, which leads to a 2.6\% uncertainty in the background estimate.

Table~\ref{tab:bkgRelErr} summarizes the systematic uncertainties and their impact on the background prediction in the SR.

The following sources of systematic uncertainty associated with the signal expectation are taken into account. The integrated luminosity is measured with precision of 2.6\%~\cite{LUM-13-001}.  Lepton trigger and identification efficiencies are measured with a precision of 2\% and 1\%, respectively.  Uncertainties in the jet energy scale and resolution correction factors yield uncertainties of 2--3\% and less than 1\%, respectively, depending on the mass hypotheses for the DM particle. Uncertainties in the b-tagging correction factors for heavy-flavour and light-flavour jets yield uncertainties of 3--4\% and less than 1\%, respectively.

\begin{table}[htb]
\centering
\topcaption{Systematic uncertainties from various sources and their impact on the total background prediction. }
\begin{tabular}{ c | c  }
\hline
\multirow{2}{*}{Source of systematic uncertainties} &  Relative uncertainty on \\
 &  total background (\%) \\ \hline
50\% normalization uncert. of other bkg in deriving SFs & 10 \\
$\mathrm{SF}_{\PW\text{+jets}}$ (CR tests) & 13  \\
\ttbar{}+jets top-quark \pt reweighting & 3.9  \\
Jet energy scale & 4.0 \\
Jet energy resolution & 3.0 \\
b-tagging correction factor (heavy flavour) & 1.0 \\
b-tagging correction factor (light flavour) & 1.8\\
Pileup model & 2.0 \\
PDF & 2.6 \\ \hline
\end{tabular}
\label{tab:bkgRelErr}
\end{table}

\section{Results}
\label{res}

Table~\ref{tab:bkg_finalcut} lists the number of events observed in the SR, along with the background prediction and expected number of signal events for a DM particle with mass of $M_{\chi}=1\GeV$ and an interaction scale $M_{*}= 100\GeV$. We observe no excess of events in the SR and set 90\% confidence level (CL) upper limits on the production cross section of DM particles in association with a pair of top quarks. The choice of 90\% CL is made in order to allow direct comparisons with related limits from astrophysical observations. A modified-frequentist $\mathrm{CL_{\mathrm{s}}}$ method~\cite{Read1,junkcls} is used to evaluate the upper limits, with both statistical and systematic uncertainties taken into account in the limit setting.

\begin{table}[htb]
\centering
\topcaption{Expected number of background events in the SR, expected number of signal events for a DM particle with the mass $M_{\chi}= 1\GeV$, assuming an interaction scale $M_{*}= 100\GeV$, and observed data. The statistical and systematic uncertainties are given on the expected yields.}
\begin{tabular}{c|c}
\hline
Source & Yield ($\pm$stat $\pm$syst) \\ \hline
$\ttbar$ &  $8.2\pm0.6\pm1.9$ \\
W & $5.2\pm1.8\pm2.1$ \\
Single top &  $2.3\pm1.1\pm1.1$ \\
Diboson & $0.5\pm0.2\pm0.2$ \\
Drell--Yan & $0.3\pm0.3\pm0.1$ \\ \hline
Total Bkg & $16.4\pm2.2\pm2.9$ \\
Data & 18  \\ \hline
\end{tabular}
\label{tab:bkg_finalcut}
\end{table}

\begin{figure}[hbt]
 \begin{center}
    \includegraphics[width=0.56\textwidth]{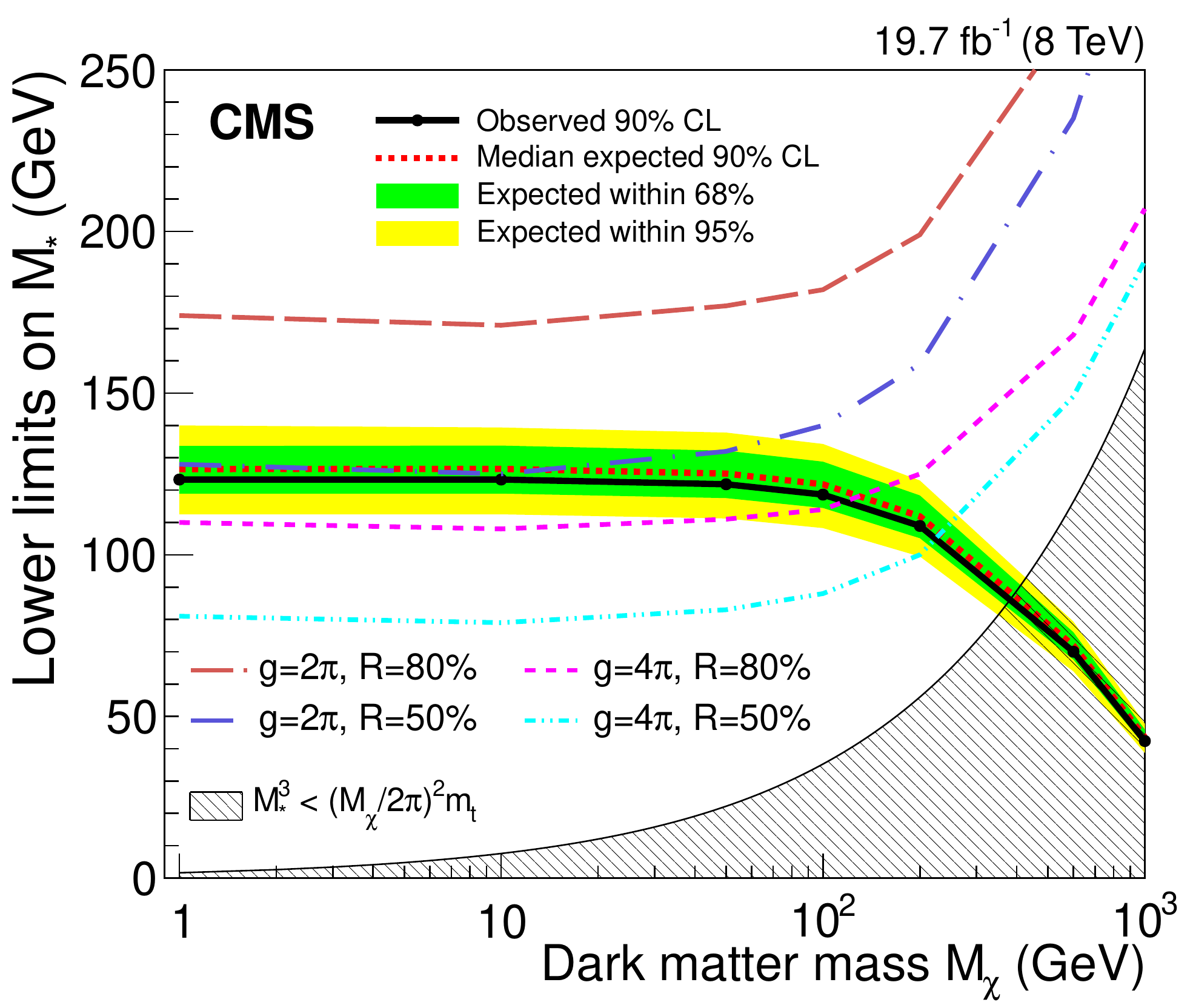}
       \caption{Observed exclusion limits in the plane of DM particle mass and interaction scale, with the region below the solid curve excluded at a 90\% CL. The background-only expectations are represented by their median (dashed line) and by the 68\% and 95\% CL bands. A lower bound of the validity of the EFT is indicated by the upper edge of the hatched area. The four curves, corresponding to different g and R values, represent the lower bound on $M_{*}$ for which 50\% and 80\% of signal events have a pair of DM particles with an invariant mass less than $g\sqrt{M^3_{*}/m_{\PQt}}$, where $g=4\pi$ and $g=2\pi$ respectively. These curves indicate further restrictions on the applicability of EFT, as explained in the text.}
 \label{fig:limit}
 \end{center}
\end{figure}

Table~\ref{tab:signal_eff} shows the signal efficiencies and the observed and expected upper limits on the $\Pp\Pp\to \ttbar+\chi\overline{\chi}$ production cross section for seven mass hypotheses of the DM particle. The relatively low values of signal efficiencies of 1--3\% are mostly due to the requirement of $\MET>320$\GeV. Cross sections larger than 20 to 55\unit{fb} are excluded at 90\% CL for DM particles with mass ranging from 1 to 1000\GeV.  Interpreting the results in the context of a scalar interaction between DM particles and top quarks, we set lower limits on the interaction scale $M_{*}$, shown in Fig.~\ref{fig:limit}. Assuming a DM particle with a mass of 100\GeV, values of the interaction scale below 118\GeV are excluded at 90\% CL.

\begin{table}[htb]
\centering
\renewcommand{\arraystretch}{1.3}
\topcaption{Expected number of signal events in SR assuming an interaction scale $M_{*}= 100\GeV$, signal efficiencies, and observed and expected limits at 90\% CL on production cross sections for $\Pp\Pp\to \ttbar+\chi\bar{\chi}$,  for various DM particle masses.}
\begin{tabular}{c|c|c|c|c}
\hline
$M_{\chi}$ (\GeVns) &Yield ($\pm$stat $\pm$syst) & Signal efficiency (\%) ($\pm$stat $\pm$syst) & $\sigma^{\lim}_{\text{exp}}$\,(fb) & $\sigma^{\lim}_{\text{obs}}$\,(fb)  \\ \hline
1 &$38.3\pm0.7\pm2.1$ & $1.01\pm0.02\pm0.05$ & $47^{+21}_{-13}$ & 55  \\
10 &$37.8\pm0.7\pm2.1$ &$1.01\pm0.02\pm0.05$ & $46^{+21}_{-13}$ & 54 \\
50 &$35.1\pm0.6\pm1.9$ & $1.20\pm0.02\pm0.06$  & $39^{+18}_{-11}$ & 45 \\
100 &$30.1\pm0.4\pm1.7$ & $1.46\pm0.02\pm0.07$ & $32^{+14}_{-9}$ & 37 \\
200 &$18.0\pm0.2\pm1.0$ & $1.73\pm0.02\pm0.08$ & $27^{+12}_{-8}$ & 32 \\
600 &$1.26\pm0.02\pm0.07$ & $2.40\pm0.03\pm0.11$ & $19^{+9}_{-6}$ & 23 \\
1000 &$0.062\pm0.001\pm0.003$ & $2.76\pm0.04\pm0.13$ & $17^{+8}_{-5}$ & 20 \\ \hline
\end{tabular}
\label{tab:signal_eff}
\end{table}

As shown in Eq.~(\ref{eft}), DM production is modeled by an EFT, an approximation that has some important limitations. Firstly, the EFT approximation is only valid when the momentum transfer $Q_{\text{tr}}$ is small compared to the mediator mass. Secondly, the couplings should not exceed the perturbative limit. Unfortunately, both of these conditions depend on the details of the unknown new physics being approximated by the EFT. For example, if we consider a model with $s$-channel exchange between the top quarks and the DM particles and a coupling equal to the perturbative limit $g\equiv\sqrt{g_{\chi}g_{\PQt}}=4\pi$, where $g_{\chi}$ and $g_{\PQt}$ are the coupling constants of the mediator to DM particles and top quarks, respectively, then we can derive a lower bound on $M_*$, $\sqrt{M^{3}_{*} /  m_{\PQt}}>M_{\chi} /  2\pi$, where $m_{\PQt}$ is the mass of the top quark~\cite{PhysRevD.82.116010,ning}. The region of parameter space in the exclusion plane that does not meet the perturbative condition for the validity of the EFT is indicated by the hatched area in Fig.~\ref{fig:limit}.

\begin{figure}[hbt]
 \begin{center}
    \includegraphics[width=0.56\textwidth]{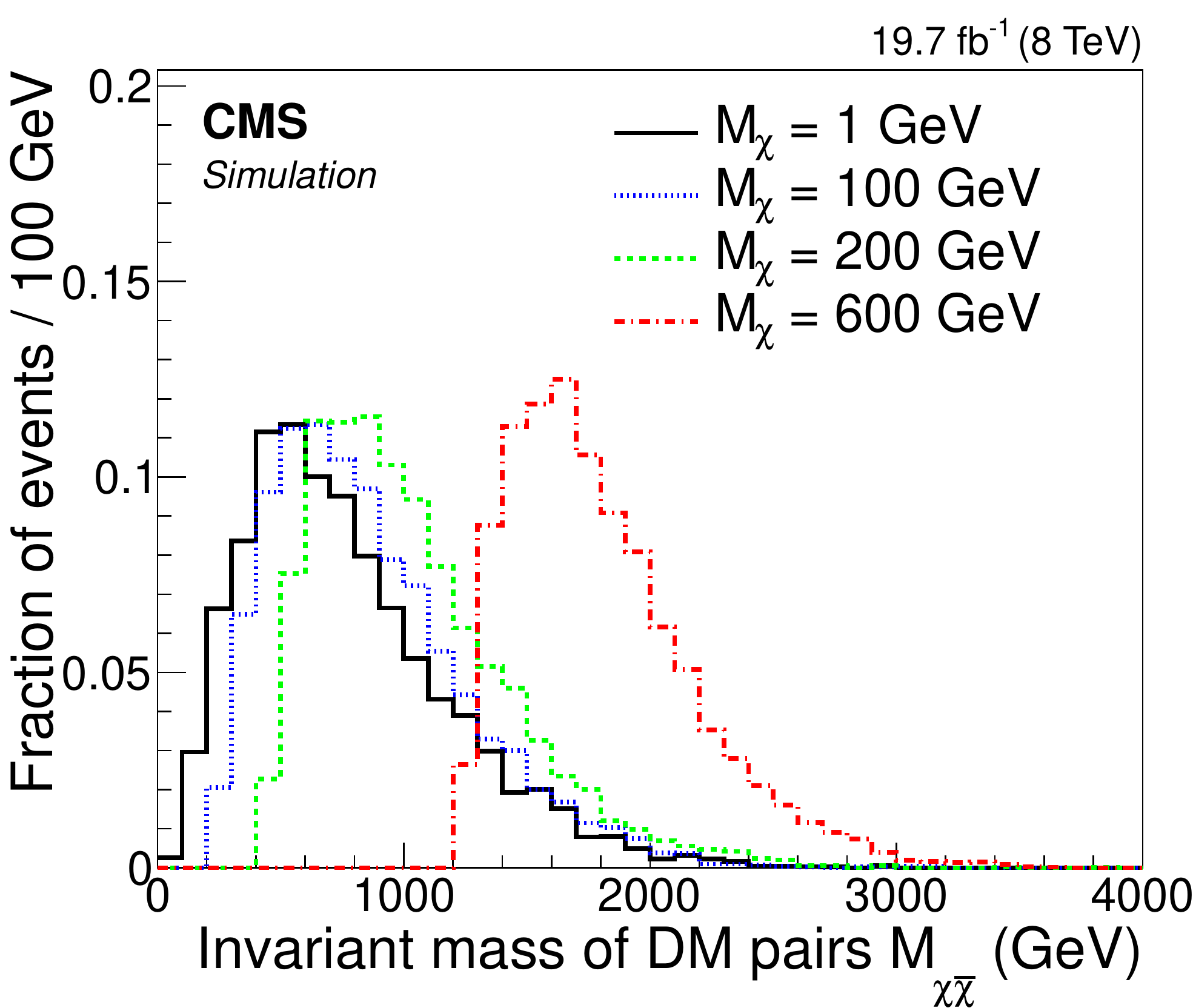}
       \caption{Invariant mass of two DM particles M$_{\chi\bar{\chi}}$ in selected signal events, for several DM mass hypotheses.}
   \label{fig:Mchichi}
 \end{center}
\end{figure}

In addition to this minimal requirement, we also test the validity of the EFT approximation with respect to the momentum transfer condition. For the same $s$-channel mediator scenario, $Q_{\text{tr}}$ is estimated as the invariant mass of two DM particles ($M_{\chi\overline{\chi}}$) as shown in Fig.~\ref{fig:Mchichi}. The EFT approximation is then valid if $M_{\chi\overline{\chi}} < g \sqrt{M^{3}_{*} /  m_{\PQt}}$. The fraction of simulated signal events that satisfy this requirement (R) is reported for given values of $g$ and $M_{*}$. For $g=4\pi$ and $g=2\pi$, contours are overlaid in Fig.~\ref{fig:limit} that indicate where in the exclusion plane 50\% or 80\% of simulated signal events passing the analysis selection criteria satisfy the momentum transfer condition. If instead of drawing such a contour we fix $M_{*}$ at the 90\% CL lower limit obtained in this analysis, then 89\% (46\%) of simulated signal events passing the analysis selection criteria satisfy the momentum requirement for $g=4\pi (2\pi)$ and $M_\chi=1$\GeV.  These fractions drop to 63\% (5\%) for $M_\chi=200$\GeV. No simulated signal events passing the analysis selection criteria are found to satisfy this requirement for $M_\chi > 600 $\GeV. For these reasons, the 90\% CL constraints on $M_{*}$ obtained in this analysis cannot be considered generally applicable, but should only be interpreted in models with large DM coupling.

\begin{figure}[hbt]
\begin{center}
    \includegraphics[width=0.56\textwidth]{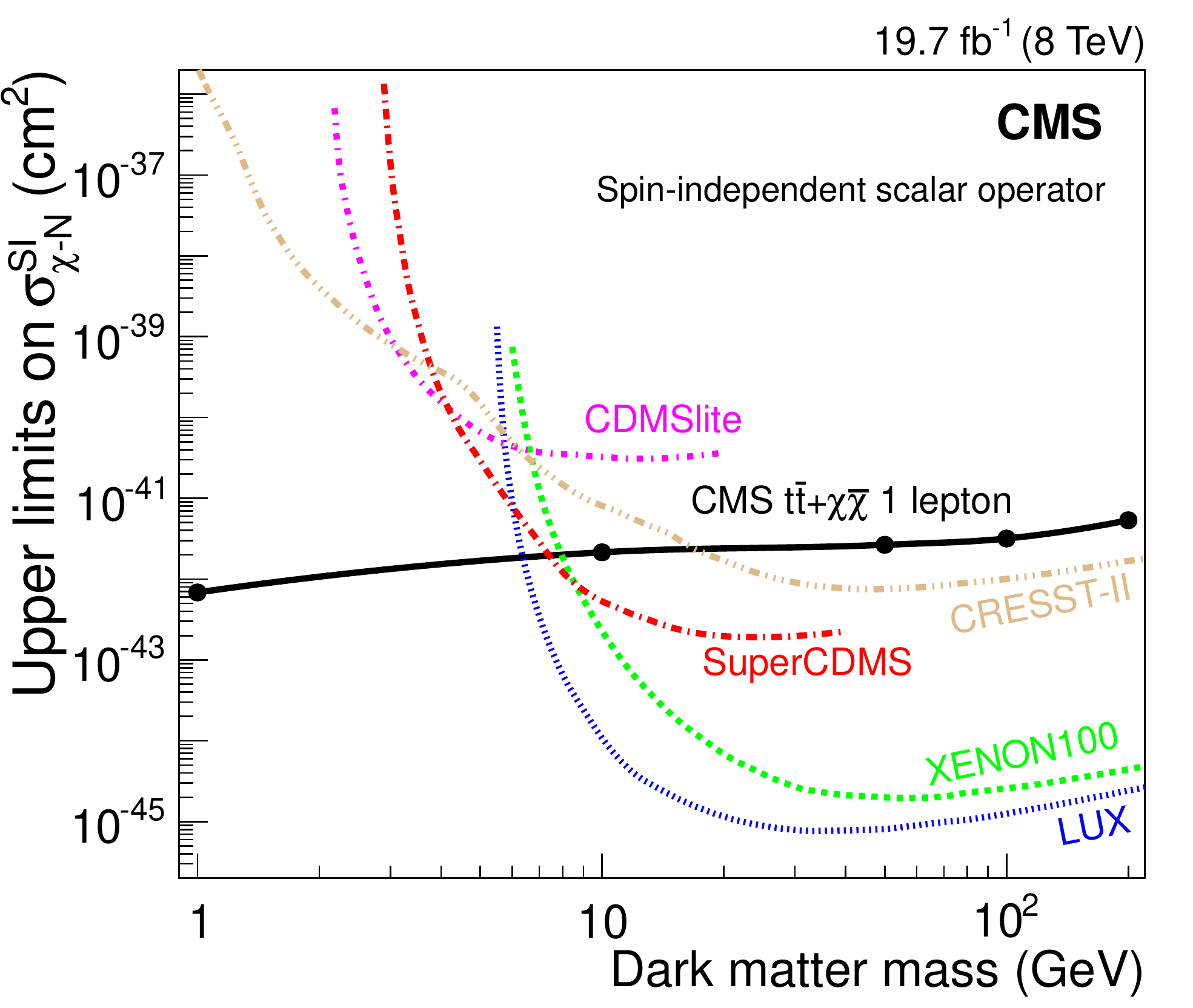}
      \caption{The 90\% CL upper limits on the DM-nucleon spin-independent scattering cross section ($\sigma^{\mathrm{SI}}_{\chi\mbox{-}\mathrm{N}}$) as a function of the DM particle mass for the scalar operator considered in this paper. Also shown are 90\% CL limits from various direct DM search experiments~\cite{PhysRevLett.112.041302,LUX,PhysRevLett.109.181301,PhysRevLett.112.241302,CRESSTII}.}
 \label{fig:limitdmxs}
 \end{center}
\end{figure}

The limits on the interaction scale $M_{*}$ can be translated to limits on the DM-nucleon scattering cross section~\cite{PhysRevD.82.116010}. Figure~\ref{fig:limitdmxs} shows the observed 90\% CL upper limits on the DM-nucleon cross section as a function of the DM mass for the scalar operator considered in this paper. More stringent limits are obtained relative to current direct DM searches in the mass region of less than $\approx$6\GeV. In this region, DM-nucleon cross sections larger than 1--$2\times10^{-42}\unit{cm}^{2}$ are excluded.

\section{Summary}\label{summary}

A search has been presented for the production of dark matter particles in association with top quarks in single-lepton events with the CMS detector at the LHC, using proton-proton collision data recorded at $\sqrt{s}= 8$\TeV and corresponding to an integrated luminosity of 19.7\fbinv. No excess of events above the SM expectation is found and cross section upper limits on this process are set. Cross sections larger than 20 to 55\unit{fb} are excluded at 90\% CL for dark matter particles with the masses ranging from 1 to 1000\GeV. Interpreting the findings in the context of a scalar interaction between dark matter particles and top quarks in the framework of an effective field theory, lower limits on the interaction scale are set. Assuming a dark matter particle with a mass of 100\GeV, values of the interaction scale below 118\GeV are excluded at 90\% CL. These limits on the interaction scale are comparable to those obtained from a similar search by the ATLAS Collaboration~\cite{ATLAS_heavyQuarks}. In the case of an $s$-channel mediator, they are only valid for large values of the coupling constant, where the effective field theory approximation holds for most signal events. These limits are interpreted as limits on the dark matter-nucleon scattering cross sections for the spin-independent scalar operator. For dark matter particles with masses below 6\GeV, more stringent limits are obtained from this search than from direct dark matter detection searches. Dark matter-nucleon cross sections larger than 1--$2\times 10^{-42}\unit{cm}^{2}$ are excluded at 90\% CL.

\begin{acknowledgments}
We congratulate our colleagues in the CERN accelerator departments for the excellent performance of the LHC and thank the technical and administrative staffs at CERN and at other CMS institutes for their contributions to the success of the CMS effort. In addition, we gratefully acknowledge the computing centres and personnel of the Worldwide LHC Computing Grid for delivering so effectively the computing infrastructure essential to our analyses. Finally, we acknowledge the enduring support for the construction and operation of the LHC and the CMS detector provided by the following funding agencies: BMWFW and FWF (Austria); FNRS and FWO (Belgium); CNPq, CAPES, FAPERJ, and FAPESP (Brazil); MES (Bulgaria); CERN; CAS, MoST, and NSFC (China); COLCIENCIAS (Colombia); MSES and CSF (Croatia); RPF (Cyprus); MoER, ERC IUT and ERDF (Estonia); Academy of Finland, MEC, and HIP (Finland); CEA and CNRS/IN2P3 (France); BMBF, DFG, and HGF (Germany); GSRT (Greece); OTKA and NIH (Hungary); DAE and DST (India); IPM (Iran); SFI (Ireland); INFN (Italy); MSIP and NRF (Republic of Korea); LAS (Lithuania); MOE and UM (Malaysia); CINVESTAV, CONACYT, SEP, and UASLP-FAI (Mexico); MBIE (New Zealand); PAEC (Pakistan); MSHE and NSC (Poland); FCT (Portugal); JINR (Dubna); MON, RosAtom, RAS and RFBR (Russia); MESTD (Serbia); SEIDI and CPAN (Spain); Swiss Funding Agencies (Switzerland); MST (Taipei); ThEPCenter, IPST, STAR and NSTDA (Thailand); TUBITAK and TAEK (Turkey); NASU and SFFR (Ukraine); STFC (United Kingdom); DOE and NSF (USA).

Individuals have received support from the Marie-Curie programme and the European Research Council and EPLANET (European Union); the Leventis Foundation; the A. P. Sloan Foundation; the Alexander von Humboldt Foundation; the Belgian Federal Science Policy Office; the Fonds pour la Formation \`a la Recherche dans l'Industrie et dans l'Agriculture (FRIA-Belgium); the Agentschap voor Innovatie door Wetenschap en Technologie (IWT-Belgium); the Ministry of Education, Youth and Sports (MEYS) of the Czech Republic; the Council of Science and Industrial Research, India; the HOMING PLUS programme of the Foundation for Polish Science, cofinanced from European Union, Regional Development Fund; the Compagnia di San Paolo (Torino); the Consorzio per la Fisica (Trieste); MIUR project 20108T4XTM (Italy); the Thalis and Aristeia programmes cofinanced by EU-ESF and the Greek NSRF; and the National Priorities Research Program by Qatar National Research Fund.
\end{acknowledgments}

\bibliography{auto_generated}

\cleardoublepage \appendix\section{The CMS Collaboration \label{app:collab}}\begin{sloppypar}\hyphenpenalty=5000\widowpenalty=500\clubpenalty=5000\textbf{Yerevan Physics Institute,  Yerevan,  Armenia}\\*[0pt]
V.~Khachatryan, A.M.~Sirunyan, A.~Tumasyan
\vskip\cmsinstskip
\textbf{Institut f\"{u}r Hochenergiephysik der OeAW,  Wien,  Austria}\\*[0pt]
W.~Adam, T.~Bergauer, M.~Dragicevic, J.~Er\"{o}, M.~Friedl, R.~Fr\"{u}hwirth\cmsAuthorMark{1}, V.M.~Ghete, C.~Hartl, N.~H\"{o}rmann, J.~Hrubec, M.~Jeitler\cmsAuthorMark{1}, W.~Kiesenhofer, V.~Kn\"{u}nz, M.~Krammer\cmsAuthorMark{1}, I.~Kr\"{a}tschmer, D.~Liko, I.~Mikulec, D.~Rabady\cmsAuthorMark{2}, B.~Rahbaran, H.~Rohringer, R.~Sch\"{o}fbeck, J.~Strauss, W.~Treberer-Treberspurg, W.~Waltenberger, C.-E.~Wulz\cmsAuthorMark{1}
% ...
\vskip\cmsinstskip
% ...
1:~~Also at Vienna University of Technology, Vienna, Austria\\
2:~~Also at CERN, European Organization for Nuclear Research, Geneva, Switzerland\\
% ...

\end{sloppypar}
\end{document}